\documentclass[graybox, natbib, nosecnum]{svmult}
\bibpunct{(}{)}{;}{a}{}{,} 

\usepackage{mathptmx}       
\usepackage{helvet}         
\usepackage{courier}        
\usepackage{type1cm}        

\usepackage{makeidx}         
\usepackage{graphicx}        
\usepackage{multicol}        
\usepackage[bottom]{footmisc}

\usepackage{color}

\makeindex

\begin{document}

\title*{Knowledge Graphs in the Libraries and Digital Humanities Domain}


\author{Bernhard Haslhofer and Antoine Isaac and Rainer Simon}

\authorrunning{Haslhofer et al.}

\institute{Bernhard Haslhofer, Rainer Simon
\at Austrian Institute of Technology, \email{firstname.lastname@ait.ac.at}
\and Antoine Isaac \at Europeana, \email{aisaac@few.vu.nl}
}

%
%
\maketitle


\section{Definition}

\textit{Knowledge graphs} represent concepts (e.g., people, places, events) and their semantic relationships. As a data structure, they underpin a digital information system, support users in resource discovery and retrieval, and are useful for navigation and visualization purposes. Within the libaries and humanities domain, knowledge graphs are typically rooted in \emph{knowledge organization systems}, which have a century-old tradition and have undergone their digital transformation with the advent of the Web and Linked Data. Being exposed to the Web, metadata and concept definitions are now forming an interconnected and decentralized global knowledge network that can be curated and enriched by community-driven editorial processes. In the future, knowledge graphs could be vehicles for formalizing and connecting findings and insights derived from the analysis of possibly large-scale corpora in the libraries and digital humanities domain.

\pagebreak

\section{Overview}

The term ``knowledge graph'' was popularized in 2012~\citep{Google:2012a} with the announcement of the ``Google Knowledge Graph'', which gathers and formalizes information from several sources in order to enhance search results. Knowledge graphs can be generic and cover a broad range of domains, or they can be tailored to a specific domain or application context. They typically integrate data from multiple, heterogeneous sources and provide both a human-interpretable representation as well as a formalized machine-readable basis for information retrieval tasks, such as (latent) semantic indexing, classification or query recommendation. Well-known examples for openly available, generic knowledge graphs are DBpedia~\citep{Auer:2007d}, Freebase~\citep{Bollacker:2008a}, or Wikidata~\citep{Vrandevcic:2014a}.

Within the libraries and digital humanities domain, the concept of a knowledge graph is deeply rooted in \textit{knowledge organization systems}, which is an umbrella term for vocabularies such as a classification schemes, thesaurus, or glossary.

Knowledge organization systems have a centuries-old tradition in the library domain, where they have been used in metadata descriptions to organize resources and facilitate their discovery and retrieval. With the advent of Linked Data~\citep{Bizer:2009a}, knowledge organization systems --- and metadata descriptions making use them --- have undergone a digital transformation and entered the realm of the World Wide Web. By linking them with semantically related Web resources within and outside the library domain, they are now forming an interconnected and decentralized global knowledge graph.

The development of knowledge graphs also has a long-standing tradition in the Humanities. For a long time, scholarly efforts have been concerned with the curation of authoritative data on e.g. places or people within a specific domain, resulting in taxonomies, gazetteers, and prosopgraphies. In terms of their digital transformation, these knowledge graphs have been following a similar pattern as knowledge organization schemes in libraries: they are increasingly being published according to Linked Data principles, and connected to other knowledge graphs on the Web, using shared semantic concepts.

Members of both domains soon realized that knowledge graphs can be curated and enriched by community-driven editorial processes, similar to those of Wikipedia. As a result, processes emerged that either allow users to directly edit a knowledge graph or provide mechanisms such as \textit{semantic tagging} to support enriching corpora with concepts from knowledge graphs.

Cross-domain knowledge graphs, as they are emerging in the libraries and digital humanities domain, open a whole new spectrum of research opportunities. They can, for instance, inform the design of quantitative analytics tasks being applied on possibly large-scale document corpora. Taking into account the numerous digitization efforts in the library and humanities domain, knowledge graphs can be vehicles for connecting and exchanging findings as well as factual knowledge gained by applying those task on copora. This can also stimulate the development of novel, mixed methods research methodologies.

\section{Knowledge Organization Systems in Libraries}

Libraries have long identified the value of maintaining (meta)data about their holdings. Over centuries the classical form of the book-based catalog has evolved into card-based catalogs and then electronic catalogs organized around the \emph{record} unit, representing data about one of the library's holdings (a monograph, a journal issue, etc). Library science developed resources and methods to rationalize the way these data were produced and exploited. The first of them were designed for classification and subject indexing, such as the Dewey Decimal Classification (DDC), the Universal Decimal Classification (UDC) or the Library of Congress Subject Headings (LCSH). Classification schemes, subject heading lists, thesauri and other taxonomies were engineered to guide the description of the subjects of library assets (or their type). Name authority files and gazetteers played a similar role as organized listings of names for persons, organizations and places. All these are composed of elements (classes, concepts, names, etc.) that are provided with lexical information (variants, synonyms) or semantic information (especially, hierarchical or associative links), which qualifies them as knowledge organization systems. They also constitute vast bodies of knowledge, with dozens of thousands of concepts (LCSH contained 342,107 authority records in April 2017) that are richly described (UDC classes exist in over 40 languages).

In parallel, the structure of records was also the subject of much rationalization efforts, with description rules being agreed at the level of thematic or geographic communities. As libraries became (early) adopters of electronic information systems, these gave rise to formats geared for exchange of bibliographic metadata, such as MARC, which defines lists of codes for the fields in records (title, author, subject, etc).

The need for structure and control within library systems came at the same time as a need to share and collaborate. Rather than each developing their own knowledge organization systems, libraries sought to share and re-use existing ones: classifications like UDC, thesauri like AGROVOC, subject heading lists like LCSH or RAMEAU are used in entire networks of libraries. They can be seen as collaboratively built systems, as all the organizations that use them can contribute updates, even as one of them plays a leading role for maintenance and quality control. A similar form of sharing happens at the level of the datasets describing assets: \emph{union cataloguing} systems have been put in place so that consortia of libraries can refer to an existing description of a book, possibly after adding to it, rather than creating a new -- probably poorer -- one from scratch. This pooling of data gives even more importance to the knowledge organization systems backing these datasets, and raised the motivation for enhancing their coverage and quality over time.

Of course, full standardization is not attainable, and overlapping vocabularies have been built, e.g., specific to a (sub-)domain or a (group of) country(ies). This became a problem when libraries needed a higher level of interoperability for their data. In some cases it has been addressed: the Virtual International Authority File (VIAF) has managed to consolidate links between the reference lists of persons and organization from dozens of (mostly national) libraries into a multilingual dataset (43 million clusters of links in 2014). MACS (Multilingual ACcess to Subjects) was another notable project, in which the national libraries of Switzerland, France and Germany have teamed with the British Library to create thousands of links across the subject heading lists LCSH, RAMEAU and SWD.

Considering the long standing knowledge organization tradition in libraries, one can conclude that libraries have established systems resembling the characteristics of today's knowledge graphs. And indeed the Mundaneum, an early 20th century attempt to build a universal repository of facts, heavily inspired by library science, has been sometimes acknowledged as a precursor of the Semantic Web notion. Yet, as the W3C Incubator Group on Library Linked Data~\citep{W3C:2011aa} has put it, more than twenty years after the invention of the Web, the library and Semantic Web communities had similar metadata concepts, but still different terminology; traditional library standards were designed only for the library community; library data was not well integrated with Web resources; and still expressed primarily in presentation-oriented, natural-language text.

\section{Library Linked Data}


The adoption of the Linked Data principles is a major step in the transition from library-centric knowledge organization systems to domain-spanning, openly available, and easily accessible knowledge graphs. These principles postulate a conceptual representation of library objects and concepts as first-class Web resources and then propose (i) the assignment of unique Web identifiers (URI) to these resources, (ii) the provision of machine-readable metadata describing these resources, and (iii) linkage of resources with semantically related resources in other datasets or knowledge organization systems. Those principles are being adopted in the publication of \textit{metadata element sets} and \textit{value vocabularies}, as detailed in the reports of the W3C Library Linked Data Incubator Group. This provided the basis for a number of data aggregation, linkage and exposure efforts producing a large variety of \textit{library linked datasets}. 


\paragraph{Metadata Element Sets}

A metadata element set provides elements (e.g., title, date, subject) to be used to describe a resource (e.g., a book), like the the aforementioned MARC. Within the scope of Linked Data, metadata element sets are typically defined using RDF Schema~\citep{Brickley:2014aa} or the OWL Web Ontology Language~\citep{Hitzler:2012aa}. Besides supporting the definition of elements (properties), those schema or ontology definition languages also provide primitives for describing groups of related resources (classes) and the relationships between these resources. Important metadata element sets within the Library Linked Data field are the Dublin Core Metadata Element Set~\citep{DCMI:2012aa}, the Bibliographic Framework Initiative (BIBFRAME) vocabulary or the Resource Description and Access (RDA) element set. Interestingly, Dublin Core was developed at the time when the notions of the Semantic Web were articulated, with several key people involved in both initiatives, which positioned it as one of the most used vocabularies in Library Linked Data and beyond.

\paragraph{Value Vocabularies}

A value vocabulary defines concepts that are used in the value space of bibliographic metadata records -- cataloging rules in libraries often require terms of certain vocabularies to be used with certain metadata elements. Value vocabularies include knowledge organization systems such as introduced above, and their semantic expressiveness ranges from flat term definition lists (glossaries), over hierarchical structures such as classification schemes and taxonomies, to more connected vocabularies like thesauri. 

Within the context of Linked Data, concepts defined as part of value vocabularies should be first-class resources identified by dereferenceable HTTP URIs. Concepts can be linked to other concepts within the same value vocabulary as well as to other concepts, possibly defined by another domain. The Simple Knowledge Organization System (SKOS)~\citep{Baker:2013aa} has become the de facto standard for expressing the basic structure and content of controlled value vocabularies.

A number of value vocabularies established in the library field were made available as Linked Data: LCSH~\citep{Summers:2008aa}, the multilingual AGROVOC thesaurus~\citep{Caracciolo:2013aa}, VIAF, several national authority files (e.g., German GND), the Getty vocabularies~\citep{Getty:2017aa} as well as parts of the Dewey Decimal Classification~\citep{OCLC:2017aa}.

\begin{figure}
  \includegraphics[width=\linewidth]{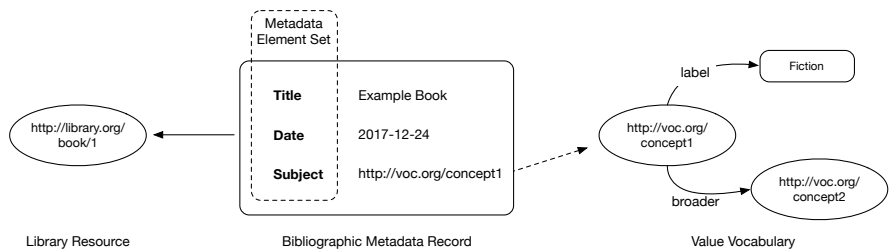}
  \caption{Library Linked Data Example and Notions.}
  \label{fig:lld_example}
\end{figure}

The terms 'ontology' and 'vocabulary' are often used to refer both to metadata element sets and value vocabularies. Their role in the making of Linked Data differ quite significantly however. Descriptions typically reuse elements from standard metadata element sets in combination with elements from value vocabularies. Figure~\ref{fig:lld_example} shows an example bibliographic metadata description with metadata fields taken from a metadata element set and a value vocabulary concept.

\paragraph{Library Linked Datasets}

Within the last decade, a number of instutions have published collections of metadata records as linked data on the Web. Such records typically describe books, but other types of cultural objects can be described too. 

The Hungarian and Swedish National~\citep{Malmsten:2008aa} libraries, who opened and exposed their OPAC catalog and corresponding authority files around the year 2008, were among the earlier adopters of the Linked Data principles. The British National Bibliography (BNB)~\citep{Deliot:2014aa} is a prominent Linked Library example which also links to external sources such as VIAF or LCSH. This was followed by many other data publication initatives on an institutional or service level, including the Amsterdam Museum~\citep{De-Boer:2012aa}, the CrossRef service, the Linked Open Library Data project, or Linked Art, which is a community working together to create a shared Linked Data model to describe art.

In parallel to instution- and service-level linked data publication initatives, major metadata aggregation hubs across libraries and other institutions, such as Europeana~\citep{Haslhofer:2011aa} and the Digital Public Library of America (DPLA)~\citep{DPLA:2017aa} started to expose collected cross-domain metadata records as Linked Open Data on a larger scale (as of 2017 Europeana and DPLA provide access to over 51 and 18 million objects, respectively) and to link them with other sources such as Wikidata. An important milestone achieved by these efforts was the wider adoption of public domain and (standard) open licenses (like the ones of Creative Commons) for the metadata. This has not only been a major business shift for libraries, but also fulfilled two important preconditions for building large-scale, cross-domain knowledge graphs: the ability to mix and match library data with data from other sources, and the possibility to edit and enhance knowledge graphs under control of a larger community.

\section{Community-driven Knowledge Graph Enrichment}

In the early days of linked data, most knowledge graphs were published out of existing institutional repositories or extracted from public knowledge sources (e.g., Wikipedia) in a rather static, 'push' fashion. The community soon realized however that knowledge graphs can be curated by community-driven editoral processes similar to those of Wikipedia.

A strong form of community-driven knowledge graph enrichment was provided by Freebase~\citep{Bollacker:2008a}, which allowed users to enhance existing and add new facts to their knowledge graph. However, Freebase was acquired by Google and shut down its public API in 2016. An even stronger form is implemented by Wikidata~\citep{Vrandevcic:2014a}, in which the user community controls not only the data and schema but also the entire editorial process. Wikidata is also important as a case of interlinking vocabularies, tackling one of the issues in traditional library data. The Wikidata community-developed alignment tool Mix'n'Match, for instance, allows users to upload a vocabulary to Wikidata and create an alignment for it, which has helped to position Wikidata as a huge hub for Library Linked Data and beyond. Near-systematic references are now available between Wikidata entities and resources from vocabularies like VIAF and GND.

Within the Digital Libraries and Digital Humanities domains, community-driven knowledge graph enrichment mostly focused on involving users in linking resources from existing corpora with concepts in knowledge graphs. One possible enrichment technique is \emph{semantic tagging}, which supports users in associating digital items or fragments thereof with concepts expressed in existing knowledge graphs. That technique has been used for semantic tagging of historical maps~\citep{Haslhofer:2013aa}, images, and texts~\citep{Simon:2015}.

\section{Knowledge Graphs in the Digital Humanities}

The development and curation of domain-specific knowledge structures has traditionally been an important element of humanities scholarship. In many cases, these structures emerge as an implicit research output, e.g., when studying the interrelation between actors and events in a specific historical setting. In other cases, the development of a knowledge organization system as such --- or components of it --- is the main objective of the research endeavor. This is the case, for example, for the development of domain taxonomies, \textit{gazetteers} or \textit{prosopgraphies} (dictionaries of people or groups of people). Initiatives such as the Tabula Imperii Romani~\citep{TIR}, the Tabula Imperii Byzantini~\citep{TIB}, the Prosopography of the Byzantine World~\citep{PBW}, the Great Britain Historical GIS~\citep{GBHGIS}, or the Treasury of Lives~\citep{ToL} have been concerned exclusively --- sometimes over a significant timespan, and long before the digital transformation ---  with the curation of authoritative data on places or people within their domain. Other efforts have focused on the translation of existing analog authority information to digital; some of them, such as the Pleiades Gazetteer of the Ancient World~\citep{Pleiades} specifically as Linked Data.

The motivation behind the development of classification schemes and authority information in the humanities is much the same as in the library domain: a need for structure, control, and a common vocabulary to facilitate collaboration. Arguably, however, scholarly humanities research differs insofar as there is a much higher degree of specificity to particular --- sometimes even niche --- domains. The heterogeneous nature of research outputs; the interpretative quality of humanities research; and the fact that work is often organized around the efforts of a single individual or small group, funded through time-limited grants, makes a “global knowledge graph of the humanities” seem an unachievable goal. Yet the need to publish data under open licenses, and build connections between datasets based on shared value vocabularies and metadata element sets is increasingly perceived in the community as key for enabling re-use; for the transparency of scholarly methods; and, ultimately, sustainability of results. This trend is reflected in the rise of Linked Data at as a theme at Digital Humanities events, conferences, and curricula, as much as in the emergence of community initiatives dedicated to establishing common interlinking standards and practices.

By and large, such initiatives advocate the idea of cross-domain linkage by means of shared  name authority files, along with recommendations on metadata element use. Crucial to this effort are, on the one hand, generic knowledge graphs - DBpedia and Wikidata in particular. On the other hand, linked data sets from libraries or museums, such as VIAF, the Getty Thesaurus of Geographic Names, or the Getty Art \& Architecture Thesaurus have been gaining traction as an interconnecting “spine” through which community-specific datasets can build outbound links to contribute to a global graph. 

As these networking efforts are enabling the aggregation of increasingly large corpora of cultural heritage content, the use of computational methods to study larger cultural phenomena is becoming more feasible (cf.~\cite{Schich2014,Cook:2012}). New tools are transforming traditional scholarship, enabling scholars to identify and and answer new research questions (cf.~\citep{Siemens:2008aa,Bodenhamer:2010,Bodard2016}). \cite{Michel:2010aa} popularized the term \textit{Culturomics} as ``the application of high-throughput data collection and analysis to the study of human culture''. In this context, knowledge organization systems represent the crucial connecting medium within an ecology of independent initiatives, that is gradually increasing mutual connectivity by creating and using Linked Open Data~\citep{Isaksen2014}.

\section{Future Directions of Research}

Large-scale knowledge graphs, as they have been emerging in the libraries and digital humanities domain, and the publication of seminal papers (e.g.,~\cite{Michel:2010aa}) demonstrate how the application of quantitive analysis to large-scale corpora opens up a spectrum of possible new research questions that, up until now, were hard to answer with existing methods and primary sources. Data was manually curated and often bound to an institution or the scope of an individual researcher's project. Most studies in the humanities and related disciplines have focused on rather small corpora, while the constitution and maintenance of institutional knowledge organization systems and related datasets in libraries required years of work for a highly skilled and trained workforce. 

Exploiting the opportunities of quantiative analysis methods poses a number of methodological, technical, and organizational challenges:  first, novel serial analysis methods and tools are needed that support scholars in in viewing, annotating, and systematically analyzing relevant parts of possibly large digitized corpora. Scholars could express relevance by selecting corresponding concept definitions in knowledge graphs. Second, scalable text-mining and machine-learning techniques are needed to systematically and efficiently analyze and compare the characteristics, contents, and relationships of concepts expressed in knowledge graphs within and across corpora. Third, algorithms are needed that support scholars in detecting, contextualizing, and analyzing various forms of expressions and associated narrative techniques in copora spanning a long time period, in which the syntax and semantics may have been subject to costant change. Under this premise, future research could focus on:

\begin{itemize}
    
    \item the development of tools and scalable techniques for aligning large scale, possible multi-media corpora with concepts expressed in knowledge graphs.

    \item the investigation of text mining algorithms that can learn from scholar's annotations and support them in investigating semantic relationships extracted from large copora.

    \item the investigation of novel reconcilation mechanisms that ensure that institutional and community-curated knowledge graphs produced in different context are truly interoperable and do not lead to ``competing'' data offers.
    
    \item the design of appropriate provenance tracking, crowd- or nichesourcing~\cite{Boer:2012aa} approaches, and validation mechanisms that ensure trust in data quality when data are curated by humans with different levels of expertise and/or result from automatic processes.

\end{itemize}

The possible research directions describe above show that there is a clear need for collaborative research among researchers from humanities, computer science as well as library information science. This will also result in novel mixed qualitative and quantiative methods for the analysis of large-scale digitized corpora relevant to the humanities and related disciplines.

\bibliographystyle{spbasic}
\bibliography{bibliography}

\end{document}